\documentclass{mn2e}

\usepackage{amsfonts}
\usepackage{amsmath}
\usepackage{graphicx}

\title[Major dry mergers at $M_* > 2\times 10^{11}M_\odot$?]
      {Evidence of major dry mergers at $M_*> 2\times 10^{11}M_\odot$ 
       from curvature in early-type galaxy scaling relations?}

\author[M. Bernardi et. al.]
{Mariangela Bernardi$^1$\thanks{E-mail: bernardm@physics.upenn.edu}, 
 Nathan Roche$^2$, Francesco Shankar$^{3}$ \& Ravi K. Sheth$^{1,4}$\\
 $^1$Department of Physics \& Astronomy, University of Pennsylvania, 
      209 S. 33rd St., Philadelphia, PA 19104, USA\\
 $^2$ Dipartimento di Astronomia, Universit\'{a} degli Studi di Bologna, 
      via Ranzani 1, I-40127 Bologna, Italy\\
 $^3$ Max-Planck-Instit\"{u}t f\"{u}r Astrophysik,
      Karl-Schwarzschild-Str. 1, D-85748, Garching, Germany \\
 $^4$ Center for Particle Cosmology, University of Pennsylvania, 
      209 S. 33rd St., Philadelphia, PA 19104, USA}
\begin{document}
\pagerange{\pageref{firstpage}--\pageref{lastpage}}

\maketitle 

\label{firstpage}

\begin{abstract}
For early-type galaxies, the correlations between stellar mass and 
size, velocity dispersion, surface brightness, color, axis ratio 
and color-gradient all indicate that two mass scales, 
$M_* = 3\times 10^{10}M_\odot$ and $M_* = 2\times 10^{11}M_\odot$, 
are special.  The smaller scale could mark the transition between 
wet and dry mergers, or it could be related to the interplay between 
SN and AGN feedback, although quantitative measures of this transition 
may be affected by morphological contamination.  At the more massive 
scale, mean axis ratios and color gradients are maximal, and above it, 
the colors are redder, the sizes larger and the velocity dispersions 
smaller than expected based on the scaling at lower $M_*$.  
In contrast, the color-$\sigma$ relation, and indeed, most scaling 
relations with $\sigma$, are not curved:  they are well-described 
by a single power law, or in some cases, are almost completely flat.  
When major dry mergers change masses, sizes, axis ratios and color 
gradients, they are expected to change the colors or velocity 
dispersions much less.  Therefore, the fact that scaling relations 
at $\sigma > 150 \, {\rm km~s}^{-1}$ show no features, whereas 
the size-$M_*$, $b/a$-$M_*$, color-$M_*$ and color gradient-$M_*$ 
relations do, suggests that $M_* = 2\times 10^{11}M_\odot$ is the scale 
above which major dry mergers dominate the assembly histories of 
early-type galaxies.  
\end{abstract}


\section{Introduction}

\begin{figure*}
 \centering
 \includegraphics[width=0.45\hsize]{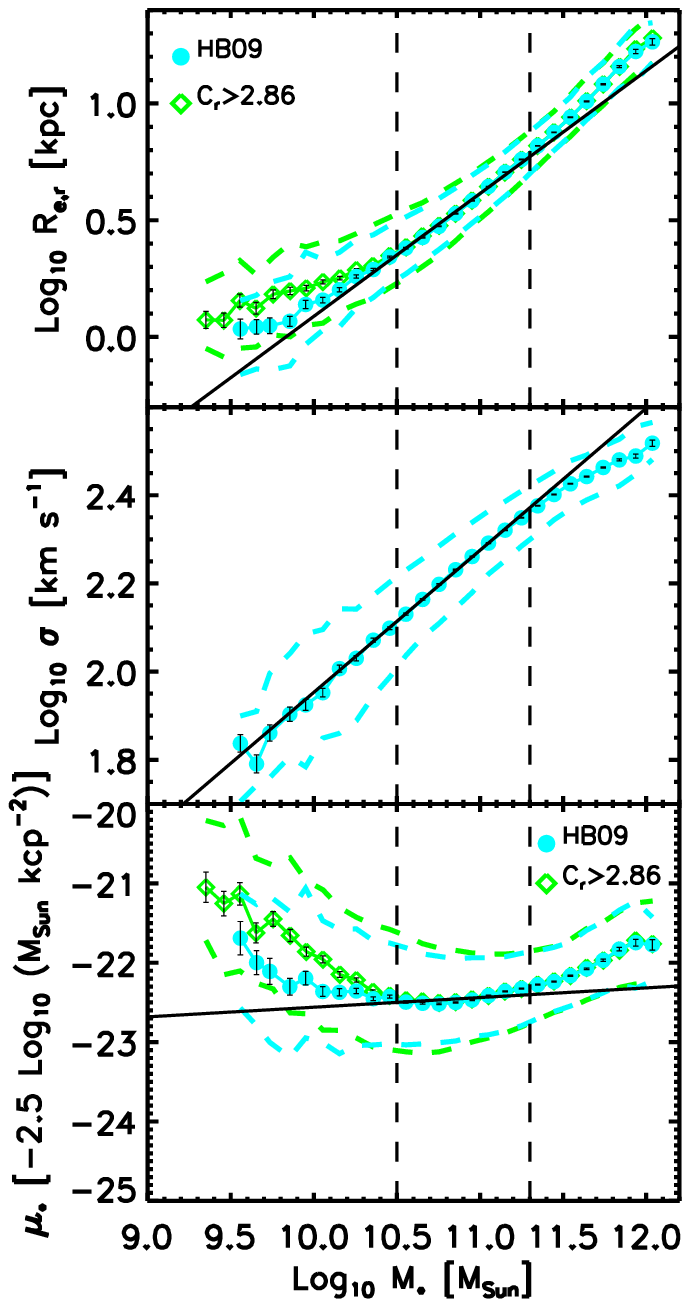}
 \includegraphics[width=0.45\hsize]{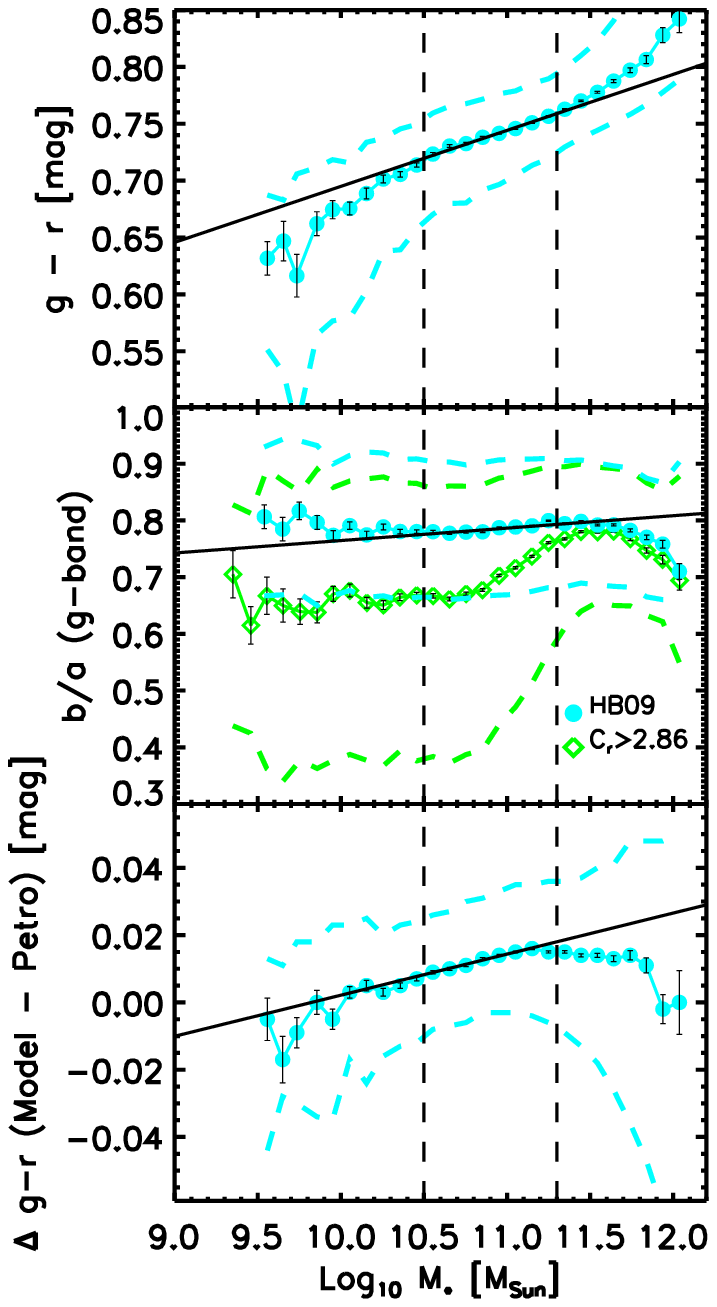}
 \caption{Curvature in the correlations between stellar mass and 
  (from top to bottom, left) size, velocity dispersion and 
  surface brightness, and color, axis ratio, and color gradient 
  (top to bottom, right), in the Hyde-Bernardi sample.  The vertical 
  dashed lines mark the scales where some of the relations change 
  slope:  $M_* = 3\times 10^{10}M_\odot$ and $2\times 10^{11}M_\odot$, 
  which correspond approximately to $M_r = -20.5$ and $-22.5$.  
  Scalings in a sample selected to have $C_r > 2.86$ are shown only 
  where they differ from the scalings in the Hyde-Bernardi sample. }
 \label{breaks}
\end{figure*}

\begin{figure*}
 \centering
 \includegraphics[width=0.45\hsize]{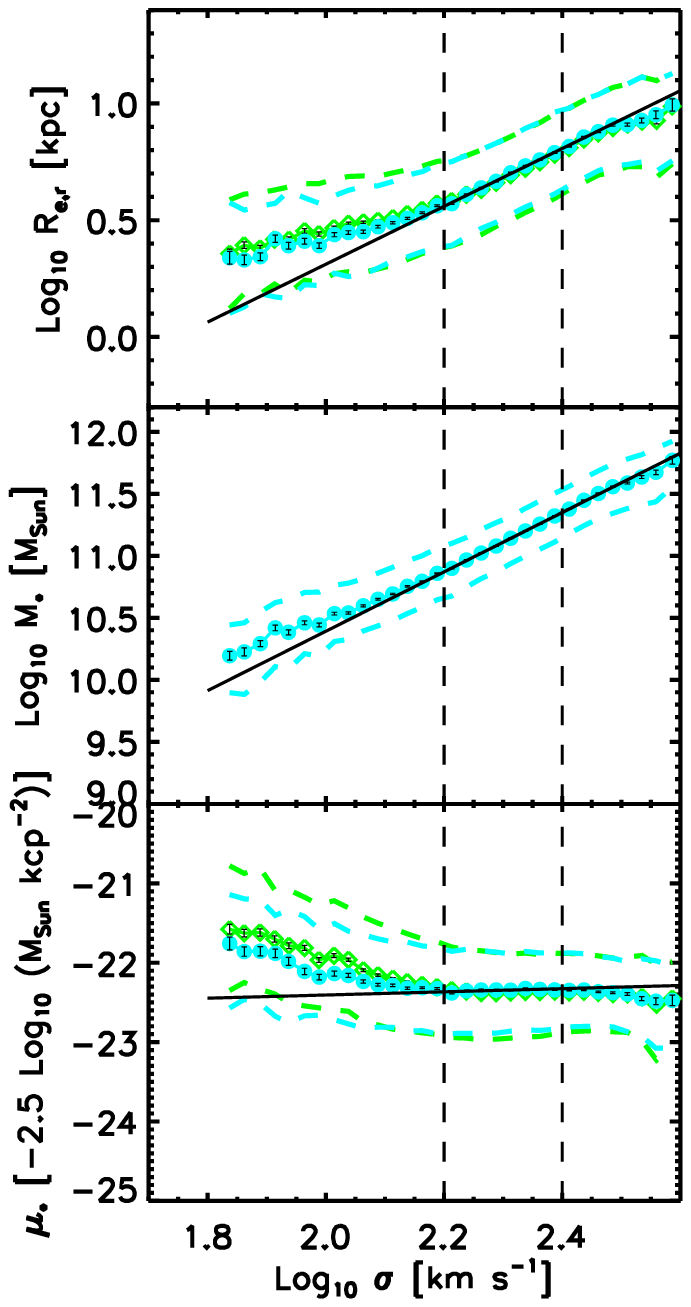}
 \includegraphics[width=0.45\hsize]{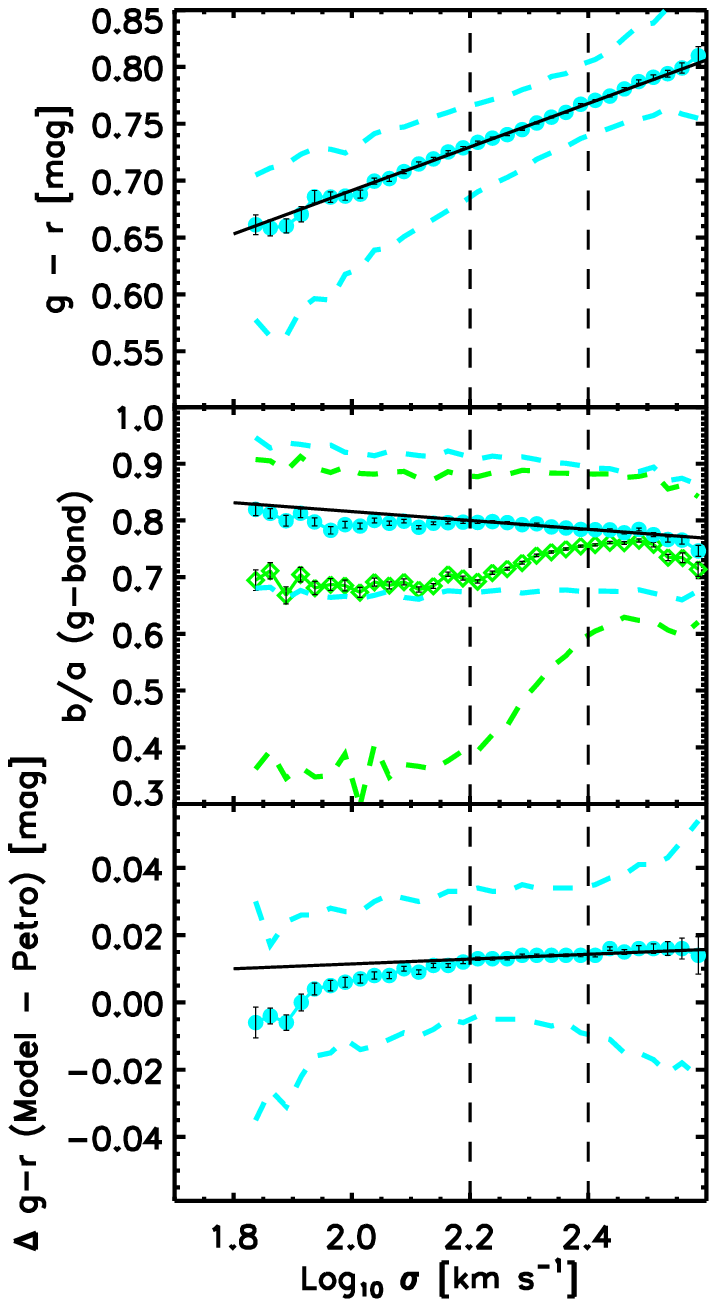}
 \caption{Curvature in the correlations between velocity dispersion 
  and (from top to bottom, left) size, stellar mass and 
  surface brightness, and color, axis ratio, and color gradient 
  (top to bottom, right), in the Hyde-Bernardi sample.  The dashed 
  lines mark the scales where one would expect to see a change in 
  the slope of the relations based on Figure~\ref{breaks}.  Scalings 
  in a sample selected to have $C_r > 2.86$ are shown only where they 
  differ from the scalings in the Hyde-Bernardi sample. }
 \label{breaksS}
\end{figure*}

Recent work (Bernardi et al. 2010b) has shown that the color-magnitude 
relation of early-type galaxies in the SDSS differs significantly from 
a pure power-law, curving downwards at low and upwards at large 
luminosities ($M_r>-20.5$ and $M_r<-22.5$, respectively).  This is 
also true of the color-size relation, and is even more apparent 
with stellar mass, where the corresponding mass scales are 
$M_* < 3\times10^{10} M_\odot$ and $M_* > 2\times 10^{11}M_\odot$, 
respectively.  The upwards curvature at the massive end does not 
appear to be due to stellar population effects.  
Curvature in the color-luminosity (or color-$M_*$) relation at the 
faint end was noticed before (e.g. Graham 2008; Skelton et al. 2009); 
the curvature at bright end is the new finding of Bernardi et al. (2010b).

Curvature at the low mass end using other parameters (e.g. surface brightness) 
has been noticed before as well (e.g. Kauffmann et al. 2003; 
Shankar et al. 2006); the subject of this Letter is to analyse non-linear 
scaling relations at the high mass end, extending the analysis of 
Hyde \& Bernardi (2009). 
We show that the curvature in the color$-M_*$ relation coincides 
with curvature in other relations with $M_*$.  However, when $M_*$ is 
replaced with velocity dispersion, then there is little curvature.  
Although curvature in scaling relations does not imply a change in the
physics which sets the relations (e.g. Graham \& Guzman 2003; Graham 2010),
we argue that our findings suggest that major dry mergers dominate the mass
 growth at $M_* > 2\times 10^{11}M_\odot$.

Our results are based on two different ways of selecting early-type 
samples from the SDSS database.  One follows Hyde \& Bernardi (2009):  
the image must be round ($b/a > 0.6$) and the light profile shape 
must be well-fit by a deVaucoleurs profile (${\tt fracDev}=1$).  
The other is a simple cut on how centrally concentrated the 
surface brightness is ($C_r > 2.86$).  The former method produces a 
sample that is more purely elliptical; the latter contains many 
edge-on disks.  See Bernardi et al. (2010a) for a more detailed 
discussion of these selection criteria, and of the SDSS photometric 
and spectroscopic parameters which we use below.  
Where necessary, we have assumed a spatially flat background cosmology 
with energy density dominated by a cosmological constant $\Lambda=0.7$, 
with a Hubble constant $H_0 = 70~{\rm km~s}^{-1}~{\rm Mpc}^{-1}$ at 
the present time.  

\section{Curvature in relations with $M_*$, but not with $\sigma$}

Figure~\ref{breaks} shows correlations between stellar mass and 
(from top to bottom, left) size, velocity dispersion and surface 
brightness, and color, axis ratio, and color gradient 
(top to bottom, right), in the Hyde-Bernardi sample.  (We 
discuss how we define the gradient in Section~\ref{sec|grad}.)
None of these correlations are pure power laws.  Although the 
curvature at $\log_{10} M_*/M_\odot < 10.5$ is interesting --
galaxies at the faint, low mass end ($M_r\ge -20.5$, 
$\log_{10}(M_*/M_\odot)\le 10.5$) tend to curve towards bluer 
colors, larger sizes, fainter surface brightnesses, smaller axis 
ratios and color gradients -- in what follows, we will focus on 
what appears to be a transition mass scale at higher masses.  
At $\log_{10} M_*/M_\odot > 11.3$, the relations curve towards larger 
sizes, smaller than expected velocity dispersions, fainter surface 
brightnesses (left panels), and smaller axis ratios and smaller 
color-gradients (right panels).  
This is precisely the mass scale on which the color-$M_*$ relation 
curves towards redder colors (top right panel).

Figure~\ref{breaksS} shows that when $M_*$ is replaced with 
velocity dispersion, then there is little curvature at 
$\log \sigma/{\rm km s}^{-1} > 2.2$.  
In fact, the correlations with surface brightness, color gradient 
and axis ratio are almost completely flat.  (That surface brightness 
and $\sigma$ are uncorrelated was noted by Bernardi et al. 2003.)  
The fact that there is no feature at the largest $\sigma$ in any of 
these relations, despite clear features in the scalings with $M_*$, 
is what has motivated this Letter.  

\begin{figure}
 \centering
 \includegraphics[width=0.98\hsize]{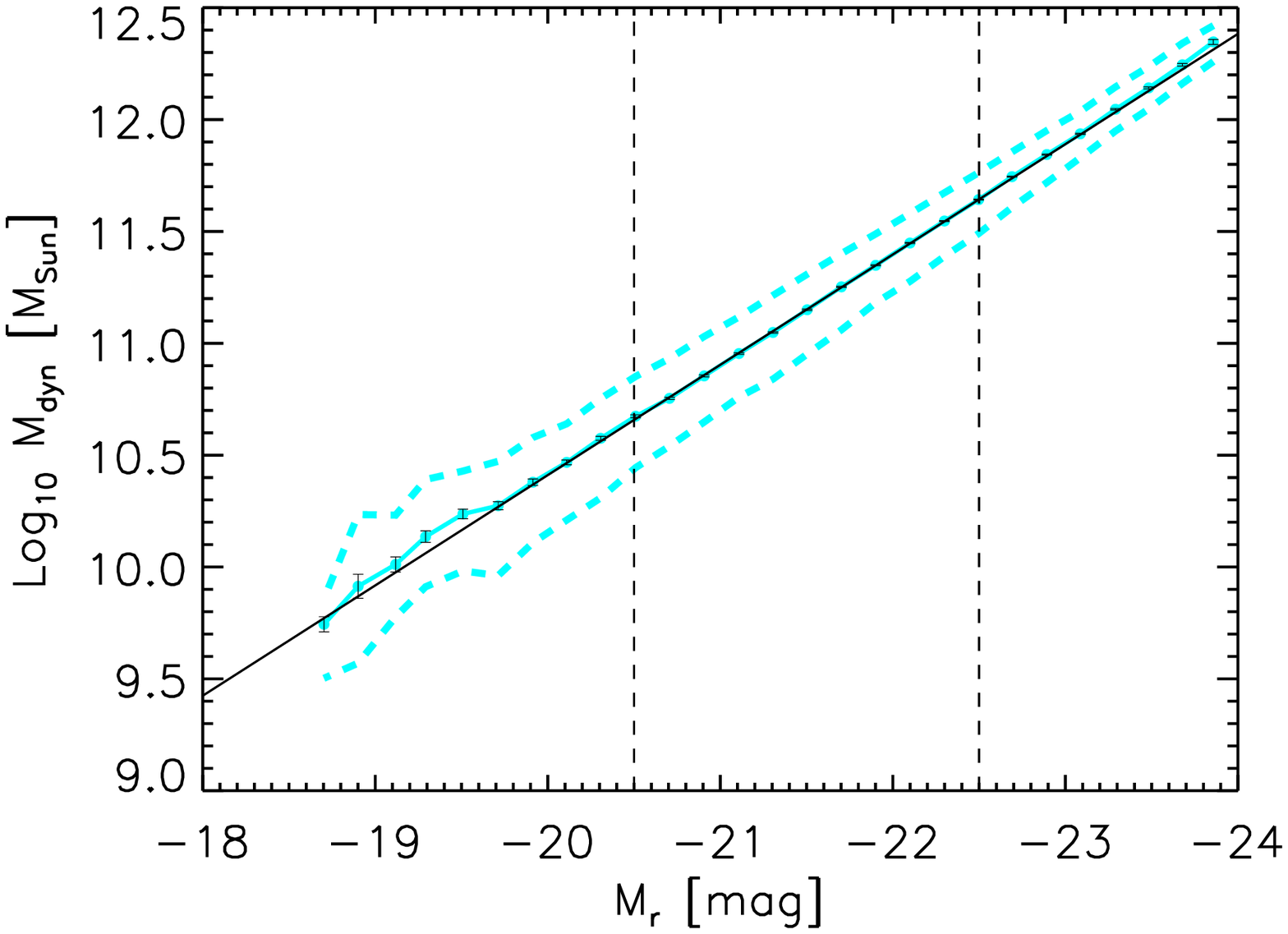}
 \includegraphics[width=0.98\hsize]{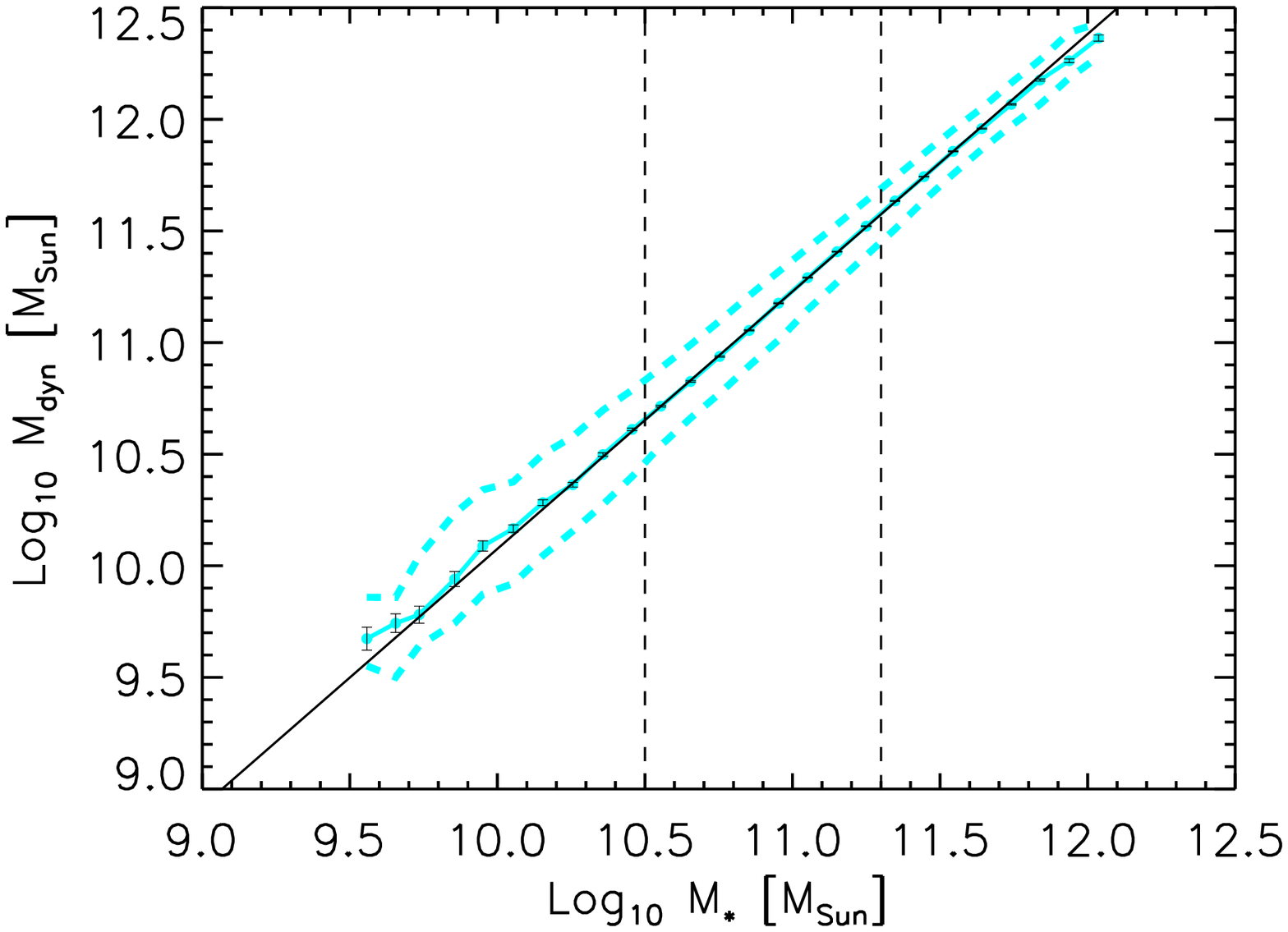}
 \caption{No curvature in the correlations between dynamical mass and 
  luminosity (top) and stellar mass (bottom).}
 \label{breaksMd}
\end{figure}

In this context, it is important to note that the relation between 
$M_{\rm dyn}\propto R\sigma^2$ and luminosity or $M_*$ is very well 
described by a single power-law over the entire range:  
the curvature in the sizes and velocity dispersions cancel 
(Figure~\ref{breaksMd}).  
Presumably, this is because the objects we observe are virialized, 
whatever their merger histories.  

\section{Discussion}
Major dissipationless mergers are expected to change the sizes in 
proportion to the masses, but to leave the velocity dispersions 
and colors unchanged.  In contrast, minor dissipationless mergers 
produce larger fractional changes in size than in mass, and decrease 
the velocity dispersions and colors (see Appendix~C in Bernardi et al. 2010b
for details).  
Therefore, the curvature in the correlations between $M_*$ and other parameters
as size, $\sigma$ and color, which have been noticed before, have all been
discussed in this context (e.g. Davies et al. 1983; Matkovic \& Guzman 2005;
Bernardi et al. 2007; Hyde \& Bernardi 2009; Bernardi et al. 2010a,b).
What is new here is the recognition that these all occur at the same
mass scale, that this mass scale is also important for axis-ratios
and color-gradients, and, significantly, that the curvature is absent
when $M_*$ is replaced by $\sigma$.

\subsection{Axis-ratios}\label{sec|ba}
The $b/a-M_*$ relation (right center panel of Figure~\ref{breaks}) 
deserves further comment.  
Van der Wel et al. (2009) report that the width of the $b/a$ distribution 
changes at $\log (M_*/M_\odot) \sim 10.5$.  They interpret this as 
evidence that, above this mass, assembly histories are dominated by 
major mergers.  Our results suggest this is not the full story.  

In Figure~\ref{breaks} we have shown two versions of this relation, 
because the Hyde-Bernardi selection requires $b/a > 0.6$.  In this 
sample, $b/a$ decreases at $\log (M_*/M_\odot) > 11.3$.  However, notice 
that this decrease is even more marked in the sample selected to have 
$C_r>2.86$, where no cut on $b/a$ is applied.  Compared to the 
Hyde-Bernardi sample, this sample has considerably smaller $b/a$ at 
small $M_*$.  Bernardi et al. (2010a) show that this is primarily due 
to an increased incidence of disks and contamination by Sas, because 
the $C_r>2.86$ sample is not as purely elliptical/early-type as the 
Hyde-Bernardi sample.  We believe this change in morphological mix is 
the primary reason why Van der Wel et al. saw what they did.  

We believe the real feature of interest is the drop in $b/a$ at 
$\log (M_*/M_\odot) > 11.3$ where (Bernardi et al. 2010a show that) 
morphological mix is no longer an issue.  Van der Wel et al. also 
see this drop, but they dismiss it.  Instead, we believe the 
narrowing of the distribution at $\log (M_*/M_\odot) \sim 10.5$ marks 
the transition from dissipational to dissipationless histories,  
or a change in relative importance of SN and AGN feedback 
(e.g. Kauffmann et al. 2003; Shankar et al. 2006), while the decrease 
in $b/a$ at $\log (M_*/M_\odot) > 11.3$ marks the transition to major 
dry mergers.   This decrease has been expected for some time 
(see Gonz{\'a}lez-Garc{\'i}a \& van Albada 2005; 
Boylan-Kolchin et al. 2006; Ragone-Figueroa \& Plionis 2007; 
Ragone-Figueroa et al. 2010) -- it was 
first found by Bernardi et al. (2008).  This is thought to indicate 
an increasing incidence of major radial mergers, since these would 
tend to result in more prolate objects.  

\subsection{Color gradients}\label{sec|grad}
The right bottom panel of Figure~\ref{breaks} shows that 
color-gradients --- here defined to be the difference between 
the {\tt model} and {\tt Petrosian} colors --- are maximal at 
$\log (M_*/M_\odot) \sim 11.3$.  (The former are approximately the 
color within the half-light radius, whereas the latter are more 
closely related to the ratio of the total luminosities in $g$ and $r$, 
so correspond to larger scales.)  
This is consistent with Figure~B1 of Bernardi et al. (2010b), 
who used the same definition of color gradient.  
It is also consistent with Roche et al. (2010), who used a different 
estimator of the gradient:  the ratio of the half-light sizes in the 
$g$ and $r$ bands.  As we discuss below, we believe that the 
appearance of this same mass scale is again signaling the onset of 
major dry mergers.  

Whereas major mergers are expected to decrease color gradients 
(e.g. Di Matteo et al. 2009), minor mergers should not change the 
gradients significantly (Kobayashi 2004) or they may enhance them 
slightly.  This is because the smaller bluer object involved in the 
minor merger is expected to deposit most of its stars at larger 
distances from the center of the object onto which it merged. 
(If it had its own gradient, then the bluest of its stars would have 
been deposited at the largest radii.)  

As a simple check of this argument, note that major mergers, which 
double $M_*$, do not change $\sigma$ (Appendix~C of 
Bernardi et al. 2010b).  Therefore, a plot of color 
gradient versus $\sigma$ should show less of a feature than when 
gradients are plotted versus $M_*$. This is indeed what we see in 
Figure~\ref{breaksS} (right bottom panel) -- the correlation is almost 
flat at $\sigma > 150 \, {\rm km~s}^{-1}$.  
Such a plot should also show greater scatter, since a range of merger 
histories, hence gradients, can all have the same $\sigma$.  While we 
do see this increase in scatter, note that the scatter in the 
gradient-$M_*$ relation grows even more dramatically -- something that 
is not easily explained.  
On the other hand, the major merger picture also provides a natural 
explanation for why none of the scaling relations in Figure~\ref{breaksS} 
show any feature at $\log\sigma/{\rm km~s}^{-1} > 2.2$, and those that 
are most clearly sensitive to merger histories are almost completely 
flat.  

Mergers are not the only way to produce or alter color gradients.  
In some models, gradients are related to feedback and winds 
(Pipino et al. 2010).  Our demonstration that gradients scale 
differently with $M_*$ than with $\sigma$ may have interesting 
implications for such models.  In addition, producing the downturn 
we see at $\log (M_*/M_\odot) > 11.3$ is an interesting challenge 
for such models, as is relating this to the changes in size, 
$\sigma$ and $b/a$ we have found.  Because the major dry merger model 
provides a simple framework for understanding all these relations, 
our results suggest that $M_* > 2\times 10^{11}M_\odot$ is the scale 
above which major dry mergers dominate the assembly history.  

\section{Implications}
We have found that a variety of early-type galaxy scaling relations -- 
the size-$M_*$, $b/a$-$M_*$, color-$M_*$ and color gradient-$M_*$ 
relations -- all show departures from a pure power-law at 
$M_* = 2\times 10^{11}M_\odot$, whereas there is no such feature when 
$M_*$ is replaced with $\sigma$.  
Since major dry mergers are expected to change the sizes, axis ratios 
and color gradients of galaxies while leaving the velocity dispersion 
and color unchanged, our findings suggest that the total stellar mass in 
early-types with $M_*>2\times 10^{11}M_\odot$ today must have grown 
primarily by relatively recent major dry mergers.  
In Bernardi et al. (2010b), we argue that such mergers may be 
required to reconcile the $z\sim 1$ counts of objects with 
$M_*>2\times 10^{11}M_\odot$ with those at $z\sim 0$.  

 This particular mass scale also appears in analyses of a local sample
 (higher quality data, but significantly smaller sample), where it is
 identified with the transition to dry mergers (see page 270
 and related discussion in Kormendy et al. 2009).
 It is special in hierarchical models also.
 Figure 3 of Guo \& White (2008) shows that below $1.6\times 10^{11}M_\odot$
 star-formation has been a significant part of the mean stellar mass
 growth rate (much of it through wet mergers),
 whereas the stellar mass growth at masses above this occurs only
 through dry mergers.  See Hopkins et al. (2008) and
 Eliche-Moral et al. (2010a,b) for other arguments suggesting dry
 mergers since $z\sim 1$ are a natural and necessary part of the
 assembly history at $M_*>2\times 10^{11}M_\odot$.  While it is reassuring
 that many lines of study all identify this same mass scale, we feel it
 worth emphasizing that our analysis suggests that above this mass
 scale, the mergers were not just dry -- they were major.
 In this respect, there is some tension between our conclusions, and 
 recent work which argues that although the mass in the central kpc 
 or so of early-type galaxies has not grown since $z\sim 2$, the half-light 
 radii have increased by more than a factor of two.  This suggests that, 
 since $z<2$, mass has been added to the outer regions only:  this sort 
 of inside-out scenario for the growth is most easily understood if the 
 mergers were minor (e.g. Lapi \& Cavaliere 2009; 
 Cook et al. 2009; Bezanson et al. 2009).  
 However, as Tiret et al. (2010) note, the observation of constant mass 
 in the central regions does not, by itself, exclude major mergers.  In 
 the simulations of Gao et al. (2004), as an object assembles its mass 
 through major dry mergers, the mix of particles in the central regions 
 can change dramatically, even though the total mass in the central regions 
 remains constant.  Our finding that color-gradients are erased at large 
 masses may be indicating that this is indeed what happens at 
 $M_*>2\times 10^{11}M_\odot$.  

Finally, it is interesting to ask how BCGs, which are amongst the most
massive objects in the local universe, fit into this picture?
Compared to non-BCGs of similar mass or luminosity, their colors are 
slightly redder (Roche et al. 2010; Figure~10 in Bernardi et al. 2010b), 
they have smaller color gradients (Roche et al. 2010), and slightly larger 
sizes (Bernardi 2009).  
Whereas the first two are in agreement with our major merger picture, 
the last suggests more size growth than is usually associated with 
major mergers.  Hence, it may be better to think of BCG formation as 
a two step process.  In the first, the major mergers which result in 
the object becoming a BCG erase its color gradient 
(and decrease $b/a$ -- Bernardi et al. 2008); thereafter, minor 
mergers puffed up its size.  Tidal stripping during the minor 
merger may also have contributed to the formation of intracluster 
light in its host halo (Bernardi 2009; Bernardi et al. 2010b).
This two step picture is in striking agreement with a detailed analysis
of the age, metallicity and abundance gradients of BCG NGC 4889
(Coccato et al. 2010).

\section*{Acknowledgments}
We are grateful to Simona Mei for help, encouragement, and for urging 
us to reorganize how we present our findings.
MB thanks Meudon Observatory, and RKS thanks the IPhT at CEA-Saclay, 
for their hospitality during the course of this work.  
MB is grateful for support provided by NASA grant ADP/NNX09AD02G; 
FS acknowledges support from the Alexander von Humboldt Foundation;  
RKS is supported in part by NSF-AST 0908241.

\label{lastpage}

\end{document}